# Maximum Power Output of Quantum Heat Engine with Energy Bath

Shengnan Liu[1], Congjie Ou[1,*]

[1] College of Information Science and Engineering, Huaqiao University, Xiamen 361021, China; 1300208003@hqu.edu.cn

* Correspondence: jcou@hqu.edu.cn; Tel.: +86-592-616-2388

**Abstract:** The difference between quantum isoenergetic process and quantum isothermal process comes from the violation of the law of equipartition of energy in the quantum regime. To reveal an important physical meaning of this fact, here we study a special type of quantum heat engine consisting of three processes: isoenergetic, isothermal and adiabatic processes. Therefore, this engine works between the energy and heat baths. Combining two engines of this kind, it is possible to realize the quantum Carnot engine. Furthermore, considering finite velocity of change of the potential shape, here an infinite square well with moving walls, the power output of the engine is discussed. It is found that the efficiency and power output are both closely dependent on the initial and final states of the quantum isothermal process. The performance of the engine cycle is shown to be optimized by control of the occupation probability of the ground state, which is determined by the temperature and the potential width. The relation between the efficiency and power output is also discussed.

# 1. Introduction

Quantum thermodynamics introduces the interdisciplinary field that combined classical thermodynamics and quantum mechanics since the concept of quantum heat engine appeared in the 1960s [1,2]. Inspired by the properties of the classical thermodynamic processes and cycles, the quantum analogues of the processes and cycles have been developed and discussed in more and more different quantum systems [3-24]. Recently, some micro sized heat engines with single Brownian particle induced by optical laser trap [25,26] and single ion held within a modified linear Paul trap [27] have been experimentally realized, which presents significant insight into the energy conversion on a microscopic level and would be expected to shed light on the experimental investigation in quantum thermodynamic characteristics of small systems. Therefore, it is of great interest to adopt a single-particle quantum system as the working substance to investigate the properties of quantum thermodynamic processes and quantum engine cycles [5,11,12,15,16,20,21,24]. A central concern of quantum thermodynamics is to understand the basic relationships between classical thermodynamics and quantum mechanics [5,13,14,15]. The quantum analog of the classical engine cycles can be set up by employing a single-particle quantum system with two energy levels [12,14,15,20] because of its simplicity.

According to the first law of thermodynamics, the quantum analogue of mechanical work and heat transfer can be defined in a natural way [5,12,14]. Thus, the basic thermodynamic processes, such as adiabatic, isochoric, isobaric ones, can be well depicted in a quantum two-state system. Nevertheless, the quantum properties of the two-state system determine the inherent difference of the thermodynamic processes. In classical thermostatistics, the law of equipartition of energy is crucial for the link between the energy and temperature [28]. However, it is violated in the quantum regime even for non-interacting particles confined in a box. In a two-state quantum system, the expectation value of the Hamiltonian depends not only on the temperature, but also on the quantum state of the system [13,14,15]. Therefore, the quantum isothermal process (to fix the temperature) and the quantum isoenergetic process (to fix the expectation value of the Hamiltonian) are totally different from each other. During the quantum isoenergetic process, the mechanical expansion/compression and the quantum state engineering are controlled simultaneously by environmental system, which is considered as energy bath [12,17,20,24]. It is worth noting that such kind of energy bath ensures the validation of the second law of thermodynamics in quantum regime [12,19,20]. Therefore, by coupling the quantum two-state system with a heat bath and an energy bath, it is possible to construct an engine cycle, which is helpful to understand the influence of quantum properties on energy conversion for a small system.

**2. Two-state Quantum System Coupled to a Heat and an Energy Bath**

The model we consider here is a single particle confined in an one-dimensional infinite square well potential with movable walls, which is a simplification of a piston. The corresponding stationary Schrödinger equation is given by $H|u_n\rangle = \varepsilon_n |u_n\rangle$ $(n=1,2,3,...)$, where $|u_n\rangle$ and $\varepsilon_n$ represent the $n$-th eigenstate and corresponding energy eigenvalue, respectively. Since we are interested in genuine quantum effects, here we assume that the temperature is low and the system size is small. In this approximation, the ground $(n=1)$ and first excited $(n=2)$ states are dominantly relevant [14,20,21]. Therefore, the occupation probabilities of the ground state and excited state can be written as $p$ and $1-p$. The expectation value of the Hamiltonian can be written as $E = p\varepsilon_1 + (1-p)\varepsilon_2$. If the system is in thermal equilibrium with the heat bath at temperature $T$, the probability of finding the system in a state with the energy $\varepsilon$ is given by the Boltzmann factor $p \propto \exp(-\varepsilon/kT)$ [3,13-15], where $k$ is the Boltzmann constant. The energy eigenvalues of the ground and first excited states are given by $\varepsilon_1 = \pi^2\hbar^2/2mL^2$ and $\varepsilon_2 = 2\pi^2\hbar^2/mL^2$, respectively, where $m$ is the mass of the particle and $L$ is the width of the square well potential. Thus, the expectation value of the Hamiltonian is

$$E = \frac{\pi^2\hbar^2(4-3p)}{2mL^2}. \qquad (1)$$

For convenience, we set $\pi^2\hbar^2/m = 1$ below. The ratio between the probabilities of the ground state and the first excited state can be written as

$$\frac{p}{1-p} = \frac{\exp(-1/2kTL^2)}{\exp(-2/kTL^2)} \qquad (2)$$

From Eq. (2) the probability that the system is in the ground state can be expressed as

$$p(T,L) = \frac{1}{1+e^{\frac{-3}{2kTL^2}}} \quad (3)$$

On the other hand, the potential width $L$ can be considered as the volume of this kind of one-dimensional system. Therefore, the force (i.e., pressure in 1 dimension) on the potential wall is [13,29],

$$f = -\left[ p\frac{d\varepsilon_1}{dL} + (1-p)\frac{d\varepsilon_2}{dL} \right] = \frac{4-3p}{L^3} \quad (4)$$

Form Eq. (4) one can see that the force varies with the potential width $L$ so it is possible to adopt a curve on the $f$-$L$ plane to describe a thermal-like quantum process. It is in fact a one dimensional analogue of the pressure-volume plane of classical thermodynamics.

If the two-state system is coupled to a thermal bath at temperature $T$, Eq. (3) can be substituted into Eq. (4) and yields

$$f = \frac{1+4e^{-\frac{3}{2kTL^2}}}{L^3(1+e^{-\frac{3}{2kTL^2}})} \quad (5)$$

According to Eq. (5), the slope of an isothermal quantum process curve on $f$-$L$ plane can be obtained as

$$\left|\left(\frac{\partial f}{\partial L}\right)_T\right| = \frac{3(4-3p)-6p(1-p)\ln[p/(1-p)]}{L^4} \quad (6)$$

If the two-state system is coupled to an energy bath to fix the expectation of Hamiltonian. From Eqs. (1) and (4) one can obtain

$$\left|\left(\frac{\partial f}{\partial L}\right)_E\right| = \frac{(4-3p)}{L^4} \quad (7)$$

Obviously, the isothermal curve on the $f$-$L$ plane is different from the isoenergetic one originating from the quantum properties.

**3. Quantum Engine Cycle Based on Two-state System**

As mentioned above, the difference between quantum isoenergetic process and quantum isothermal process can be illustrated by their curves on the $f$-$L$ plane. According to the quantum adiabatic theorem [5,30-32], which should not be confused with the thermodynamic adiabaticity, if the time scale of the change of the Hamiltonian or the potential width is much larger than the typical dynamical one, $\hbar/E$, then the stationary Schrödinger equation for the energy eigenstate holds instantaneously [17].

The slope of the curve of the adiabatic process, during which the state remains unchanged (i.e., $p$ is fixed) can be directly derived from Eq. (4) as follows:

$$\left|\left(\frac{\partial f}{\partial L}\right)_p\right| = \frac{3(4-3p)}{L^4} \tag{8}$$

It is worth noting that for the positive temperature, $T > 0$, Eq. (3) indicate that $1/2 < p < 1$. In this case, the slopes of quantum isothermal process, isoenergetic process and adiabatic process can be compared at the same potential width $L$ and yields [29],

$$\left|\left(\frac{\partial f}{\partial L}\right)_p\right| > \left|\left(\frac{\partial f}{\partial L}\right)_T\right| > \left|\left(\frac{\partial f}{\partial L}\right)_E\right|, \quad (1/2 < p < 1) \tag{9}$$

According to Eq. (9), if the positive temperature area is concerned, we can construct the possible three-process cycles on the *f-L* plane as it is shown in Figure 1.

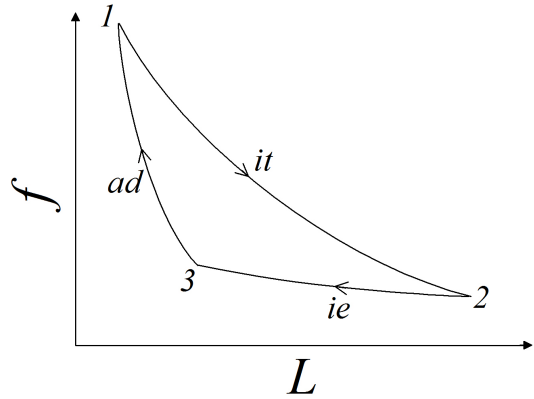

**Figure 1.** The diagram of the constructed quantum cycle on the *f-L* plane, where *ie*, *ad* and *it* represent the isoenergetic, adiabatic and isothermal quantum processes, respectively.

$1 \to 2$ is an isothermal quantum process, the system is coupled to a heat bath with temperature $T_H$. During the expansion of the potential width, one wall of the potential acts as a piston to perform work [17] and the energy is transferred from the heat bath to the system. $2 \to 3$ is an isoenergetic quantum process, which means that the two-state system exchanges energy with an energy bath to keep its expectation value of the Hamiltonian constant. $3 \to 1$ is an adiabatic quantum process to connect the first two processes so that a closed cycle on the *f-L* plane can be realized.

**4. Performance of The Quantum Engine Cycle**

During the isothermal process $1 \to 2$, the heat absorbed from the heat bath is,

$$Q_{in} = T_H \Delta S = T_H (S_2 - S_1), \tag{10}$$

where *S* is the entropy of the two-state system and it is given by,

$$S_i = k\left[ p_i \ln \frac{1-p_i}{p_i} - \ln(1-p_i) \right] \quad (i=1,2), \tag{11}$$

where $p_i$ is the occupation probability of the ground state when the system is at point "*i*" of the *f-L* plane. Substitution of Eq. (3) into Eq. (11) yields

$$S_i = \frac{3(1-p_i)}{2T_i L_i^2} - k \ln p_i \quad (i=1,2) \tag{12}$$

Since points 1 and 2 are connected by an isothermal process with temperature $T_H$ in the *f*-L plane, one has $T_1 = T_2 = T_H$ in Eq. (12). Therefore, Eq. (10) can be rewritten as,

$$Q_{in} = T_H(S_2 - S_1) = \frac{3}{2}\left[ \frac{1-p_2}{L_2^2} - \frac{1-p_1}{L_1^2} \right] - kT_H \ln \frac{p_2}{p_1} \tag{13}$$

During the isoenergetic compression process $2 \to 3$, the expectation of Hamiltonian is fixed. From the first law of thermodynamic [5,32], the heat released from the system to the surroundings is compensated by the work, i.e.,

$$Q_{out} = W_{23} = \int_2^3 f_{23} dL = \frac{4-3p_2}{L_2^2} \ln \frac{L_3}{L_2} < 0 \tag{14}$$

On the other hand, during the adiabatic process $3 \to 1$, the quantum state is fixed (i.e., no transitions between the states), that is, $d'Q = 0$. Therefore, the work performed during one full cycle is $W_{tot} = Q_{in} + Q_{out}$ and accordingly the efficiency of the cycle can be obtained as

$$\eta = \frac{W_{tot}}{Q_{in}} = 1 - \frac{\frac{4-3p_2}{L_2^2}\ln\frac{L_2}{L_3}}{T_H(S_2 - S_1)} \quad (15)$$

From Eq. (3) one can also obtain

$$T_i L_i^2 = \frac{3}{2k \ln \frac{p_i}{1-p_i}} \quad (i=1,2,3) \quad (16)$$

During the isoenergetic process $2 \to 3$, one has $E = (4-3p_2)/2L_2^2 = (4-3p_3)/2L_3^2$ to yield

$$\frac{L_2^2}{L_3^2} = \frac{4-3p_2}{4-3p_3} \quad (17)$$

Substituting Eq. (17) into Eq. (15), and considering that the quantum state is fixed during the adiabatic process $3 \to 1$ (i.e., $p_3 = p_1$), one can have,

$$\eta = 1 - \frac{k(4-3p_2)\ln\frac{p_2}{1-p_2}\ln\frac{4-3p_2}{4-3p_1}}{3(S_2 - S_1)} \quad (18)$$

From Eq. (18) one can see that the efficiency of such three-process quantum engine cycle depends on $p_1$ and $p_2$. It means that the properties of quantum state are crucial for performance of the quantum engine of this kind. In the classical point of view, the efficiency of engine cycle is described in terms of the thermodynamic variables, such as pressure, temperature, volume, etc., whereas the concept of quantum states is also relevant in the quantum regime. In fact, the probabilities of ground states, $p_i$, are functions of temperature $T_i$ and volume $L_i$, as indicated in Eq. (3). By this relationship, we can also analyze the behavior of Carnot efficiency in a similar way.

From Eq. (16) one can have

$$T = \frac{3}{2kL^2 \ln \frac{4-2EL^2}{2EL^2 - 1}} \quad (19)$$

and consequently obtain the variation of temperature with respect of potential width during the isoenergetic process [29],

$$\left(\frac{\partial T}{\partial L}\right)_E = \frac{3}{kL^3}\left(\frac{1}{\ln\frac{4-2EL^2}{2EL^2-1}}\right)^2 \left(\frac{6EL^2}{(4-2EL^2)(2EL^2-1)} - \ln\frac{4-2EL^2}{2EL^2-1}\right) > 0 \qquad (20)$$

During the isoenergetic compression process, from Eq. (20) one can easily find that the temperature decreases with the compression of the potential width. On the other hand, during the adiabatic compression process $3 \to 1$, the probability distribution of each energy level is fixed. From Eq. (3) one can obtain $TL^2 = const$, which means that the temperature increases with the decreasing of potential width. Therefore, the lowest temperature $T_C$ is at point 3 on the f-$L$ plane and the highest temperature $T_H$ is at the isothermal process $1 \to 2$. Suppose that there is a quantum Carnot cycle composed by two quantum isothermal processes and two quantum adiabatic processes, working between $T_H$ and $T_C$. The efficiency of it coincides with the classical Carnot cycle [15], say,

$$\eta_C = 1 - \frac{T_C}{T_H} \qquad (21)$$

By using Eqs. (16) and (17), the quantum Carnot efficiency can be rewritten as,

$$\eta_C = 1 - \frac{(4-3p_2)\ln\frac{p_2}{1-p_2}}{(4-3p_1)\ln\frac{p_1}{1-p_1}} \qquad (22)$$

Eqs. (18) and (22) are both the functions of $p_1$ and $p_2$. Therefore, we can compare $\eta$ with $\eta_C$ by varying $p_1$ and $p_2$. It is worth noting that from Eq. (3) one can obtain,

$$\left(\frac{\partial p}{\partial L}\right)_T = \frac{2}{L} p(1-p)\ln\frac{1-p}{p} \qquad (23)$$

Eq. (23) shows that $(\partial p/\partial L)_T < 0$ when the positive temperature is considered, i.e., $1/2 < p < 1$. It means that the probability of find the system in the ground state of the two-state system decreases during the isothermal expansion, which indicates $p_1 > p_2$. Therefore, the 3D plot of $\eta$ and $\eta_C$ varying with $p_1$ and $p_2$ can be shown in Figure 2.

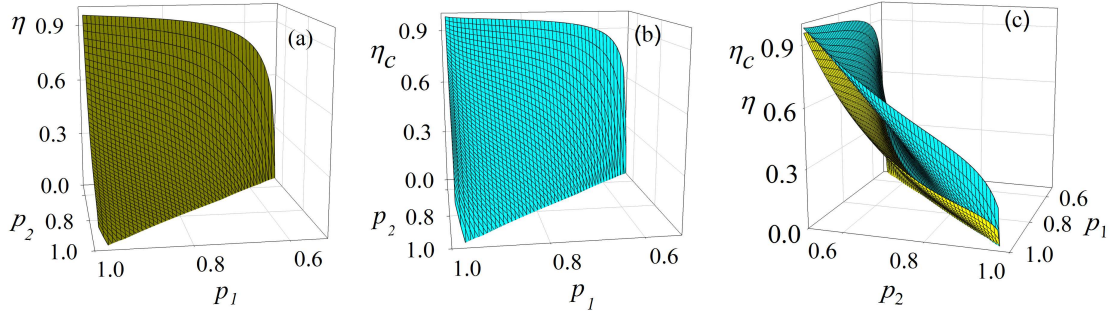

**Figure 2.** Comparison between $\eta$ and $\eta_C$ at the positive temperature region, $1/2 < p_2 < p_1 < 1$ where $p_1$ and $p_2$ are ground state probabilities of the two-state system at points 1 and 2 in *f-L* plane, respectively. (a) $\eta$ varies with $p_1$ and $p_2$; (b) $\eta_C$ varies with $p_1$ and $p_2$; (c) the combination of (a) and (b).

From Figure 2 one can see that for every possible pair of $p_1$ and $p_2$, $\eta$ is always smaller than $\eta_C$, as expected. It is worth noting that in our previous work [29], another 3-process quantum engine cycle was constructed by following sequence: "isoenergetic process → adiabatic process → isothermal process". There exist a non-monotonic relationship between efficiency and $\Delta T \equiv (T_H - T_C)$ when $T_H$ is larger than the characteristic value of temperature $T_{H,C}(E)$. However, in the cycle described by Fig. 1, the non-monotonic relationship disappears. In fact, the cycle in Fig. 1 and the one in Ref. [29] are two separate parts of a quantum Carnot cycle [15], as shown in Fig. 3. According to Eq. (1), the expectation value of the Hamiltonian depends on potential width $L$ and ground state probability $p$. It is possible to find a set of $(p_2, L_2, p_3, L_3)$ that satisfy

$$\frac{(4-3p_2)}{2L_2^2} = \frac{(4-3p_3)}{2L_3^2} \qquad (24)$$

which means that the expectation value of the Hamiltonian at point 2 equals to that of point 3. Therefore, points 2 and 3 can be connected by an isoenergetic quantum process on the *f-L* plane.

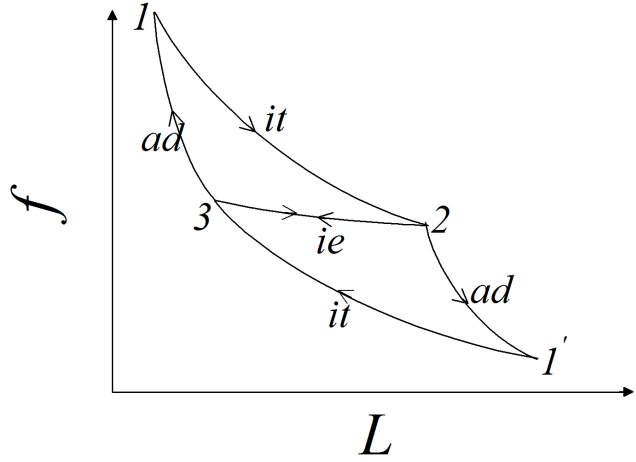

**Figure 3.** A quantum Carnot cycle is composed of two isothermal processes ($1 \to 2$ and $1' \to 3$) and two quantum adiabatic processes ($2 \to 1'$ and $3 \to 1$). It is a quantum isoenergetic process that connects points 2 and 3. "$1 \to 2 \to 3 \to 1$" cycle is identical to Fig. 1 and "$3 \to 2 \to 1' \to 3$" is another kind of 3-process cycle discussed in Ref. [29].

The efficiency of cycle "$3 \to 2 \to 1' \to 3$" in Fig. 3 is given in [29],

$$\eta' = 1 - \frac{3(S_2 - S_3)}{k(4-3p_3)\ln\frac{p_3}{1-p_3}\ln\frac{4-3p_2}{4-3p_3}} \tag{25}$$

Since from point 3 to point 1 is a quantum adiabatic compression process, the quantum state of the two-state system does not change. Therefore, one can have $p_3 = p_1$ as well as $S_3 = S_1$ and then Eq. (25) can be rewritten as,

$$\eta' = 1 - \frac{3(S_2 - S_1)}{k(4-3p_1)\ln\frac{p_1}{1-p_1}\ln\frac{4-3p_2}{4-3p_1}} \tag{26}$$

From Eqs. (18), (22) and (26) one can verify the following relationship,

$$\eta_C = \eta + (1-\eta)\eta' \tag{27}$$

Eq. (27) shows clearly that Carnot efficiency can be precisely reproduced by ideal coupling of the two 3-process cycles indicated in Fig. 3. We stress that, in the classical Carnot cycle, it is not possible to connect point 2 and 3 by a thermodynamic process because of the absence of the isoenergetic process.

It shows again that the 3-process quantum cycle discussed above has no counterpart in classical thermodynamics.

Inspired by the finite-time thermodynamics [17], we can discuss the power output of the above mentioned 3-process quantum engine cycle. As indicated in Fig. 1, the potential wall moves from point 1 to point 2 and then moves back after one full cycle and the total movement of it can be expressed as $2(L_2 - L_1)$. Assuming that this velocity is small in order to avoid transition to higher excited states, but still with finite average speed $\bar{v}$. The total cycle time can be expressed as $\tau = 2(L_2 - L_1)/\bar{v}$. Therefore, the power output is given by,

$$P = \frac{Q_{in} + Q_{out}}{\tau} = \frac{\left\{\frac{3}{2}\left[\frac{1-p_2}{L_2^2} - \frac{1-p_1}{L_1^2}\right] - kT_H \ln\frac{p_2}{p_1} + \frac{4-3p_2}{L_2^2}\ln\frac{L_3}{L_2}\right\}\bar{v}}{2(L_2 - L_1)} \quad (28)$$

Substituting Eqs. (16), (17) and (24) into Eq. (28) yields,

$$P = \frac{3\left[(1-p_2)\ln\frac{p_2}{1-p_2} - (1-p_1)\ln\frac{p_1}{1-p_1}\right] - 3\ln\frac{p_2}{p_1} + (4-3p_2)\ln\frac{p_2}{1-p_2}\ln\frac{4-3p_1}{4-3p_2}}{4\frac{L_1^3}{\bar{v}}\ln\frac{p_1}{1-p_1}\left(\left(\frac{\ln\frac{p_1}{1-p_1}}{\ln\frac{p_2}{1-p_2}}\right)^{1/2} - 1\right)} \quad (29)$$

Eq. (29) indicates that the power output is a function of $p_1$ and $p_2$ if the initial potential width $L_1$ and average speed $\bar{v}$ are given. For the sake of convenience, we discuss the behavior of dimensionless power output, $P^* = PL_1^3/\bar{v}$, below. With the positive temperature condition, $1/2 < p_2 < p_1 < 1$, the variation of $P^*$ with $p_1$ and $p_2$ can be shown in Fig. 4.

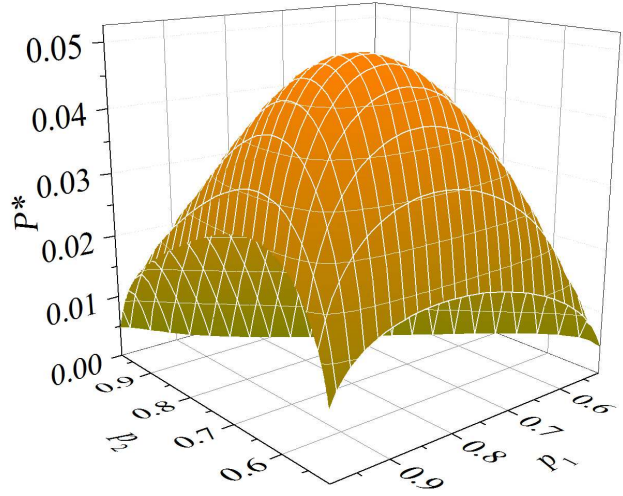

**Figure 4.** Dimensionless power output $P^*$ with respect of $p_1$ and $p_2$

From Fig. 4 one can find that there exist a global maximum value for $P^*$. More precisely, $P^*_{max}$ can be obtained by solving the following coupled equations,

$$\begin{cases} \dfrac{\partial P^*}{\partial p_1} = 0 \\ \dfrac{\partial P^*}{\partial p_2} = 0 \end{cases} \quad (30)$$

The numerical result shows that $P^*_{max} = 0.052$ when $p_1 = 0.86$ and $p_2 = 0.62$. Thus, the power output can be optimized by adjusting the probabilities of ground states at point 1 and 2 on the *f-L* plane. From Fig. 4 it can also be seen that for any given value of $p_1$, the curve of $P^*$ versus $p_2$ is always concave to give the global maximum. From Eq. (16) we can see that a given $p_1$ indicates a given temperature $T_H$ if the potential width at the initial point is set. During the expansion process $1 \rightarrow 2$, the system is coupled to a heat bath with temperature $T_H$, i.e.,

$$T_H = \dfrac{3}{2kL_1 \ln \dfrac{p_1}{1-p_1}} = \dfrac{3}{2kL_2 \ln \dfrac{p_2}{1-p_2}} \quad (31)$$

Eq. (31) shows that $L_2$ will tend to infinite if $p_2$ is close to $1/2$, which indicates that a full cycle time will be very large and yields zero power output. On the other hand, if $p_2$ is very close to $p_1$, the area of cycle $1 \rightarrow 2 \rightarrow 3 \rightarrow 1$ on the *f-L* plane tends to zero. Vanishing work also means zero power output. Therefore, the power output can be optimized in the region $1/2 < p_2 < p_1$.

Furthermore, Eqs. (18) and (29) show that the efficiency and power output are both functions of $p_1$ and $p_2$. Therefore, we can generate the curves of power output with respect to the efficiency by varying $p_1$ and $p_2$ under the condition $p_2 < p_1$. Fig. 5 shows the $P^*$ vs. $\eta$ relationship for some values of $p_1$.

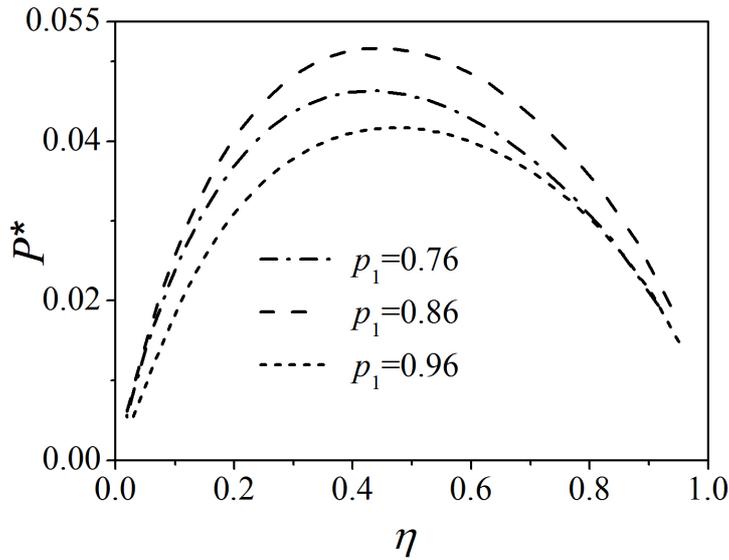

**Figure 5.** Dimensionless power output $P^*$ versus efficiency $\eta$ for some given values of $p_1$

From Fig. 5, one can find that all the $P^*$ vs. $\eta$ curves are concave. So there exists an efficiency $\eta^*(p_1)$ that corresponding to the maximum power output $P^*_{max}(p_1)$ for each value of $p_1$. The physical meaning of each $\eta^*(p_1)$ is nontrivial. When $0 < \eta < \eta^*$, the power output increases with the increasing of efficiency. It means that the cycle is not working in optimal regions. Both efficiency and power output can be optimized towards positive direction. When $\eta^* < \eta < 1$, the power output is decreasing with the increasing of $\eta$. It means that in order to improve the engine's efficiency, the cost is to decrease the engine's power output, and vice versa. Therefore, this kind of trade-off between the efficiency and power output should be concerned when the engine is working at this region, and $\eta^*$ is the lower bound of the region.

## 5. Conclusions

With the analysis of a two-state quantum particle trapped in an infinite square well, a 3-process quantum cycle was proposed by coupling the system to a heat bath and an energy bath, respectively. Based on the difference between isothermal process and isoenergetic process in quantum thermodynamics, the heat transferred into quantum cycle and total work performed during one cycle were obtained to yield the efficiency $\eta$. Comparison between $\eta$ and Carnot efficiency $\eta_C$ showed that the quantum Carnot cycle can be constructed by the combination of two symmetrical 3-process quantum cycles, in spite of the fact that the isoenergetic quantum process has no counterpart in classical thermodynamics. Furthermore, by considering the average speed of square potential wall, the power output of this kind of 3-process cycle was shown. It was found that the probability distributions at the starting and ending points of the isothermal expansion process are crucial to optimize the cycle performances. It was also shown that there exists a region of preferable performance, where the efficiency is still high and the power output is not low. These features of the present engine may suggest experiments of a new kind.

**Acknowledgments:** Project supported by the Natural Science Foundations of Fujian Province (Grant No. 2015J01016), the Program for prominent young Talents in Fujian Province University (Grant No. JA12001), Program for New Century Excellent Talents in Fujian Province University (Grant No. 2014FJ-NCET-ZR04), Scientific Research Foundation for the Returned Overseas Chinese Scholars (Grant No. 2010-1561), and Promotion Program for Young and Middle-aged Teacher in Science and Technology Research of Huaqiao University (Grant No. ZQN-PY114).

**Authors' contributions:** C.O. conceived the idea, formulated the theory. S.L. and C.O. designed the model, carried out the research. S.L. and C.O. wrote the paper.

**Competing Interests:** The authors declare that they have no competing interests.